\documentclass{pasj00}
\draft

\begin{document}
\SetRunningHead{Author(s) in page-head}{Running Head}
\Received{2009/12/31}%{yyyy/mm/dd}
\Accepted{2009/01/01}%{yyyy/mm/dd}

\title{Suzaku Observation of the Diffuse X-Ray Emission from the Open
Cluster Westerlund~2: a Hypernova Remnant?}

%%% begin:list of authors
% Do NOT capitalize all letters in "textsc".
\author{Yutaka \textsc{Fujita}\altaffilmark{1},
Kiyoshi \textsc{Hayashida}\altaffilmark{1},
Hiroaki \textsc{Takahashi}\altaffilmark{1},
and
Fumio \textsc{Takahara}\altaffilmark{1},
}
\altaffiltext{1}{Department of Earth and Space Science, Graduate School of
Science, \\Osaka University, Toyonaka, Osaka 560-0043}
\email{fujita@vega.ess.sci.osaka-u.ac.jp}
%%% end:list of authors

%%% Please use the following style in case that sorting by 
%%% affilation is impossible. 
%
% \author{%
%   D-Firstname \textsc{D-Familyname}\altaffilmark{1}
%   E-Firstname \textsc{E-Familyname}\altaffilmark{1,2}
%   and
%   F-Firstname \textsc{F-Familyname}\altaffilmark{2}}
% \altaffiltext{1}{Address of Institute}
% \email{ddddd@xxx.xxx.xx.xx}
% \email{eeeee@xxx.xxx.xx.xx}
% \altaffiltext{2}{Address of Institute}

%% `\KeyWords{}' always has to be placed before `\maketitle'.
\KeyWords{stars: winds, outflows --- ISM: cosmic rays --- ISM:
individual (RCW~49) --- ISM: supernova remnants --- Galaxy: open
clusters and associations: individual (Westerlund~2)} 
%Do NOT move this preamble from here!

\maketitle

\begin{abstract}
We present the analysis of Suzaku observations of the young open cluster
Westerlund 2, which is filled with diffuse X-ray emission. We found that
the emission consists of three thermal components or two thermal and one
non-thermal components. The upper limit of the energy flux of the
non-thermal component is smaller than that in the TeV band observed with
H.E.S.S. This may indicate that active particle acceleration has stopped
in this cluster, and that the accelerated electrons have already
cooled. The gamma-ray emission observed with H.E.S.S. is likely to come
from high-energy protons, which hardly cool in contrast with electrons.
Metal abundances of the diffuse X-ray gas may indicate the explosion of
a massive star in the past.
\end{abstract}

\section{Introduction}

Cosmic-ray particles with energies of $\lesssim 10^{15}$~eV are thought
to be accelerated in our Milky Way. Supernova remnants (SNRs) have long
been considered to be the main sources of the galactic cosmic-rays.  In
fact, their synchrotron emission reveals the existence of high-energy
electrons in SNRs \citep{koy95}, and their 
gamma-ray emission may
indicate that of high-energy protons \citep{aha04}. However, it is not
clear to what extent the isolated SNRs could account for the cosmic-ray
particles in our Milky Way.

Young open clusters are one of other candidates of the cosmic-ray
sources.  The HEGRA stereoscopic system of air \v{C}erenkov telescopes
has detected a gamma-ray signal inside the core of the OB association
Cygnus OB2 \citep{aha02}. The Milagro gamma-ray observatory found
gamma-ray emission around the young open cluster Berkeley 87
\citep{abd07}. These observations suggest that particles are accelerated
around open clusters.

Several models have been proposed as the origin of the gamma-ray
emission from young open clusters. The gamma-rays could be produced
through collisions of protons accelerated at shocks in massive stellar
winds in the cluster \citep{gio96}. Some of the protons may penetrate
the winds of the massive stars toward the stellar surfaces and produce
gamma-rays through hadronic interactions in the dense parts of the winds
\citep{tor04}. High-energy leptons may contribute to the gamma-ray
emission through inverse Compton scattering \citep{bed07,man07}. In
addition to the stellar winds, pulsars in a cluster might be responsible
for the gamma-ray emission \citep{bed03}. The gamma-rays could also be
produced through the photodisintegration of highly boosted nuclei
followed by daughter deexcitation \citep{anc07}.

Westerlund~2 is one of the clusters from which gamma-ray emission has
been detected with H.E.S.S. \citep{aha07}. It is ionizing the large
H\emissiontype{II} region RCW 49 (NGC 3247). The gamma-ray emission is
extended ($\sim 0.2\degree$), and the whole cluster is buried in it.
Observations in the X-ray band are crucial to find the origin of the
diffuse gamma-ray emission from the cluster and the mechanism of
particle acceleration. In particular, the strength of non-thermal X-ray
emission can be used to discriminate between the hadronic and the
leptonic origin of the gamma-rays \citep{bed07}, and estimate the
time-scale of particle acceleration \citep{man07}. In the X-ray band,
although the cluster has been observed with Einstein \citep{her84,gol87}
and ROSAT \citep{bel94}, strict constraint on non-thermal emission has
not been obtained. With Chandra, \citet{tow05} studied diffuse X-ray
emission $\lesssim 5'$ from the center of Westerlund~2 and found that
the spectrum could be represented by both thermal and non-thermal
emission.

In this paper, we report the results from Suzaku observations of
Westerlund~2. Since Suzaku has a large collecting area and low
background \citep{mit07}, it is the best instrument for observations of
dim and diffuse X-ray emission. This paper is organized as follows. The
observation is presented in section~\ref{sec:obs}. In
section~\ref{sec:spec}, we explain the details of the spectral
analysis. Section~\ref{sec:dicuss} is devoted to discussion. Conclusions
are summarized in section~\ref{sec:conc}.  We assume that the distance
to the cluster is 5.4~kpc unless otherwise mentioned \citep{fur09},
although there is a debate about it ($\sim 2$--8~kpc;
\cite{pia98,dam07,tsu07,rau07,asc07}). At a distance of 5.4~kpc, $1'$
corresponds to 1.6~pc.

\section{Observations}
\label{sec:obs}

\begin{figure}
  \begin{center}
    \FigureFile(80mm,80mm){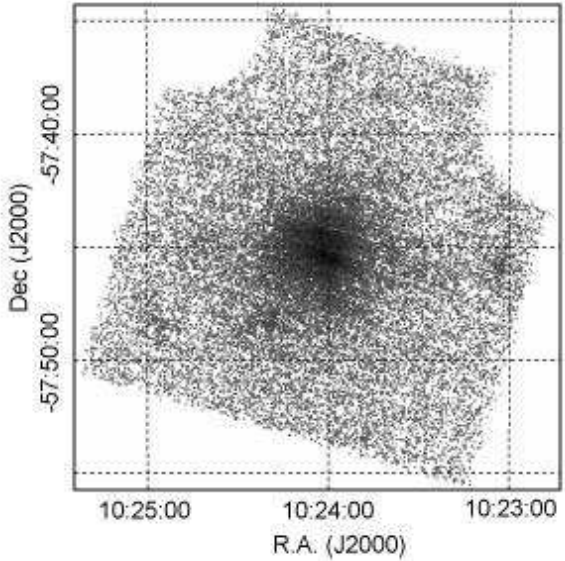}
    %%% \FigureFile(width,height){filename}
  \end{center}
  \caption{XIS~0 raw image of Westerlund~2. The image is not corrected
 for vignetting and background. The calibration sources at the corner of
 the field were excluded.}\label{fig:xis_im}
\end{figure}

Westerlund~2 was observed with Suzaku on 2008 August 9--11 and 2009
February 4--5. The exposure times for the two observations are 73.7~ks
and 33.5~ks, respectively. The data of the two observations are dealt
together in the following analysis. Suzaku has three working XIS CCDs
\citep{koy07}. Two of them (XIS~0, and 3) are front-illuminated (FI) and
one (XIS~1) is back-illuminated (BI). The XIS was operated in the normal
full-frame clocking mode. The edit mode was $3\times 3$ and $5\times 5$
and we combined the data of both modes for our analysis. We employed the
following calibration files: XIS (20090203) and XRT (20080709). Events
with ASCA grades of 0, 2, 3, 4, and 6 were retained. We excluded the
data obtained at the South Atlantic Anomaly, during Earth occultation,
and at the low elevation angles from Earth rim of $<5\degree$. We
further excluded the data with elevation angle from the rim of the
shining Earth smaller than $20\degree$. Figure.~\ref{fig:xis_im} is a raw
XIS image of Westerlund~2.

\section{Spectral Analysis}
\label{sec:spec}

We analyze the spectrum of the diffuse emission around Westerlund~2.
Since the angular resolution of the XIS is moderate ($\sim 2'$), we need
to estimate the amount of X-ray emission from point sources that
contaminate the diffuse X-ray emission from the cluster (e.g.
\cite{ham07}; \cite{hyo08}; \cite{ezo09}). Moreover, since the diffuse
emission is faint, the background spectrum must be constructed
carefully. Errors on fitted spectral parameters are given at the 90\%
confidence level from now on.

\subsection{Contamination from Point Sources}

In order to estimate the contamination of the point sources, we used
Chandra archive data of Westerlund~2 (ID 3501, 6410, and 6411). In order
to extract the positions and spectra of the point sources, we used the
ACIS Extract (AE) software package\footnote{The ACIS Extract software
package and User's Guide are available online at
http://www.astro.psu.edu/xray/docs/TARA/ae\_users\_guide.html.}
\citep{Broos02}. The procedures used in AE are described in
\citet{TFM03} and \citet{Getman05b}. AE detected 930 point sources from
the data of the three observations ($\gtrsim 10^{-15}\:\rm ergs\:
s^{-1}\: cm^{-2}$ in the 2--10~keV band). AE produces spectral files for
individual point sources. For the 9 brightest sources with photon counts
of $>300$, we can treat their spectra separately. For fainter point
sources, it is difficult to obtain their individual spectra because of
their small photon counts. Thus, we combined their spectral files by
{\tt MATHPHA} included in FTOOLS. For that purpose, we used the data of
570 point sources, which are well-separated from their neighboring
sources, and their background spectra can be obtained around them.  We
assume that the spectral shapes of the individual fainter sources are
the same as that of the combined spectrum but their luminosities are
different.

\begin{figure}
  \begin{center}
    \FigureFile(80mm,80mm){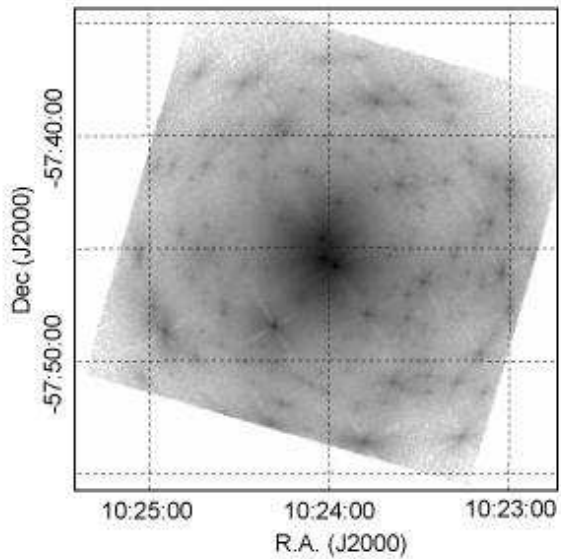}
    %%% \FigureFile(width,height){filename}
  \end{center}
  \caption{Simulated XIS~0 image of Westerlund~2. Only photons from
 point sources are included. The exposure time is assumed to be 22~Ms in
 order to emphasize the X-ray structure.}\label{fig:xis_point}
\end{figure}

Using the positional and spectral data, we simulate an XIS observation
of the point sources with the XIS simulator {\tt MKPHLIST} and {\tt
XISSIM} \citep{ish07}. The produced event files include both positional
and energy information of photons only from the point
sources. Figure~\ref{fig:xis_point} is the simulated XIS image of
Westerlund~2 that includes only photons from the point sources.

\subsection{Background Spectrum}
\label{sec:bg}

Since the TeV gamma-ray emission from Westerlund~2 covers the entire XIS
field \citep{aha07}, we used the spectrum of a blank region around
SWIFTJ1010.1--5747 obtained with Suzaku (\timeform{10h10m55s.4},
\timeform{-57D51m14s} [J2000.0]) as the background of Westerlund~2. The
exposure time is 38~ks. We call this region the blank region. This
region is $\sim 1.7\degree$ away from Westerlund~2
(\timeform{10h23m59s.0}, \timeform{-57d44m40s}). The Galactic
coordinates of Westerlund~2 and the blank region are $(l,
b)=(284.25\degree, -0.32\degree)$ and $(282.87\degree, -1.38\degree)$,
respectively. Westerlund~2 and the blank region are both affected by the
Galactic Ridge X-ray Emission (GRXE; \cite{rev06}).

We subtract the non-X-ray instrumental background (NXB) from the
spectrum of the blank region. The NXB is constructed from night Earth
data, and is generated by the routine {\tt XISNXBGEN} included FTOOLS,
considering the time-variation of the NXB \citep{taw08}. We calculate
the effective area for the Suzaku XIS chips using {\tt XISSIMARFGEN},
which provides the ancillary response file (ARF) through Monte Carlo
simulations \citep{ish07}. After the NXB is subtracted and removing a
few noticeable point sources brighter than $\sim 10^{-13}\:\rm ergs\:
s^{-1}\: cm^{-2}$ in the 2--10~keV band, we fit the X-ray emission from
the blank region with spectral components of the GRXE, the cosmic X-ray
background (CXB), and the local hot bubble (LHB). We assume that the
contribution of faint point sources ($\gtrsim 3\times 10^{-15}\:\rm
ergs\: s^{-1}\: cm^{-2}$ in the 2--10~keV band), which Chandra could
have detected with an exposure time of $\sim 100$~ks, to the Suzaku
spectrum of the blank region can be ignored, because \citet{ebi05} found
that the contribution of point sources brighter than $\sim 3\times
10^{-15}\:\rm ergs\: s^{-1}\: cm^{-2}$ (2--10~keV) to the GRXE is only
$\sim 10$~\% (see also \cite{rev09}). The spectrum of the GRXE can be
represented by two thermal models (VAPEC), and two absorptions (PHABS)
with a combination of $\rm PHABS*VAPEC+PHABS*VAPEC$. The spectra of the
CXB and the LHB are represented by an absorbed power-law model ($\rm
PHABS*POWER$) and a thermal model (APEC), respectively. For the CXB
spectrum, we use the results of \citet{kus02} for POWER and those of
\citet{dic90} and \citet{kal05} for PHABS. Since the region around
Westerlund~2 is affected by the GRXE, we use the average of the LHB
spectra at positions $\pm 5~\degree$ away from Westerlund~2 in the
direction of galactic latitude \citep{sno98}. The temperature of the LHB
is 0.1~keV. The parameters for the GRXE in the blank region are shown in
Table~\ref{tab:blank}. The best-fit values for the GRXE are derived by
fixing all the parameters of the spectra of the CXB and the LHB at their
best-fit values. For the groupings of metals, we refer to the results in
\citet{ebi05}. For the errors, we include not only statistical errors
but also systematic errors made by a spatial variation of the intensity
of the CXB. \citet{kus02} estimated the spatial variation of the CXB is
$6.49$~\% for the ASCA field. We adjusted the value to be consistent
with the XIS field. Moreover, since we masked the regions around bright
point sources ($\sim 20$\% of the XIS field), we consider the decrease
of the effective area in the field. Thus, we adopt the systematic error
of 16\% for the CXB flux.

\begin{table}
  \caption{Best-fit parameters for the GRXE}\label{tab:blank}
  \begin{center}
    \begin{tabular}{llc}
  \hline              
Component & Parameter & \\ 
      \hline
Soft  & $k T$ (keV)   
                  & $0.41_{-0.04}^{+0.28}$  \\
& Abundance except for Ne, Mg, Si (solar) 
                  & 0.044$^*$               \\
& Ne (solar)      & $0.22_{-0.06}^{+0.07}$  \\
& Mg (solar)      & $0.11_{-0.08}^{+0.09}$  \\
& Si (solar)      & $0.25_{-0.25}^{+0.35}$  \\
& $N_H$ ($10^{22}\rm\: cm^{-2}$)  
                  & $0.16_{-0.16}^{+0.13}$  \\
& Observed flux$^\dagger$ ($10^{-8}\:\rm ergs~cm^{-2}~s^{-1}~sr^{-1}$)
                  & $1.7_{-0.0}^{+0.1}$     \\
& Intrinsic flux$^\dagger$ ($10^{-8}\:\rm ergs~cm^{-2}~s^{-1}~sr^{-1}$)
                  & $2.8_{-0.2}^{+0.1}$     \\
\hline 
Hard  & $k T$ (keV)   
                  & $8.9_{-2.4}^{+4.3}$     \\
& Abundance except for Fe (solar) 
                  & 0.17$^*$                \\
& Fe (solar)      & $1.5_{-0.6}^{+1.1}$     \\
& $N_H$ ($10^{22}\rm\: cm^{-2}$)  
                  & $0.76_{-0.41}^{+0.48}$  \\
& Observed flux$^\dagger$ ($10^{-8}\:\rm ergs~cm^{-2}~s^{-1}~sr^{-1}$)
                  & $6.3_{-0.7}^{+0.9}$     \\
& Intrinsic flux$^\dagger$ ($10^{-8}\:\rm ergs~cm^{-2}~s^{-1}~sr^{-1}$)
                  & $8.0_{-1.2}^{+1.2}$     \\
\hline
& $\chi^2$/d.o.f. & 353/331                 \\   
\hline 
    \end{tabular}
  \end{center}
\hspace{30mm}$^*$ Fixed at the values of the GRXE in Table~8 of
\citet{ebi05}.

\hspace{30mm}$^\dagger$ In the 0.7--10~keV band.
\end{table}

Considering the distance between Westerlund~2 and the blank region on
the sky, we need to take account of spatial variation of the GRXE, when
we use the spectrum of the blank region as the background spectrum of
Westerlund~2. In the 3--20~keV band, the flux of the GRXE in the blank
region is $2.0\pm 0.3\times 10^{-11} \rm\: ergs\: cm^{-2}\: s^{-1}\:
deg^{-2}$ and is consistent with the value obtained with RXTE ($2.3\pm
0.4\times 10^{-11} \rm\: ergs\: cm^{-2}\: s^{-1}\: deg^{-2}$;
\cite{rev06}). At the position of Westerlund~2, the X-ray flux of the
GRXE is $1.8\pm 0.4\times 10^{-11} \rm\: ergs\: cm^{-2}\: s^{-1}\:
deg^{-2}$ \citep{rev06}, which is $\sim 60$\% of that of the CXB. Thus,
we change the normalization of the GRXE spectrum of the blank region
accordingly, considering the systematic error of the flux. Moreover, we
include the systematic error of the CXB flux. Since the area of the
masked regions in the Westerlund~2 field is different from that in the
blank region (see the next subsection), we adopt an error of 18\%.

\subsection{Diffuse Emission from Westerlund~2}
\label{sec:diffuse}

Including the leaked X-rays from the point sources and the X-ray
background emission (GRXE, CXB, and LHB) estimated above, we analyze the
XIS spectrum of the diffuse emission from Westerlund~2. We refer to the
diffuse X-ray emission around Westerlund~2 excluding those unwanted
components as the gas component. We calculate the effective area of the
Suzaku XIS chips using {\tt XISSIMARFGEN}. All spectra were rebinned to
give a minimum of 20 raw counts per spectral bin to allow $\chi^2$
statistics to be applied. The NXB spectrum calculated with {\tt
XISNXBGEN} is subtracted from the spectrum. In order to minimize the
contamination from the point sources, we mask the central region of the
cluster ($<2'$ from the center). We also mask 88 bright point sources
with circles of $0.5'$--$2'$. In total, we mask $\sim 40$\% of the XIS
field.

In the spectral fits, FI and BI XIS spectra were respectively summed. We
fix the parameters for the leaked stellar emission and the X-ray
background except for the normalizations of the GRXE and the CXB
spectra, which we allow to vary according to the systematic errors
estimated in section~\ref{sec:bg}. We find that the spectrum of the gas
component can be represented by three thermal models or two thermal
models and one power-law model with absorptions. That is, $\rm
PHABS*VAPEC+PHABS*(VAPEC+VAPEC)$ or $\rm
PHABS*VAPEC+PHABS*(VAPEC+POWER)$. We call the former 3T model and the
latter 2TP model. In each model, metal abundances are the same among
thermal models with different temperatures. The abundances of N and O
are linked to that of C, and the abundance of Ni is linked to that of
Fe. The abundances of He, Al, Ar, and Ca are fixed at the solar
values. The parameters of these models are common between the two types
of XISs (FI and BI), except for the normalization of the highest
temperature VAPEC model (3T) or the power-law model (2TP), which is the
main component of the spectra.

\begin{figure}
  \begin{center}
    \FigureFile(80mm,80mm){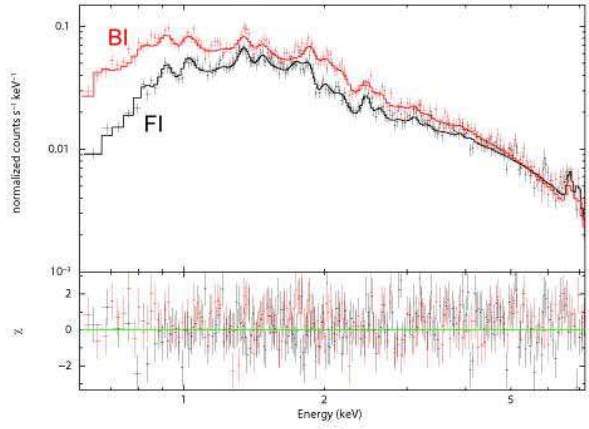}
    %%% \FigureFile(width,height){filename}
  \end{center}
  \caption{X-ray spectrum of the diffuse emission from Westerlund~2
(crosses). The NXB spectrum is subtracted. The result of the fit is
shown by the lines, while the lower panel plots the residuals divided by
the 1~$\sigma$ errors.}\label{fig:xis}
\end{figure}

\begin{figure}
  \begin{center}
    \FigureFile(80mm,80mm){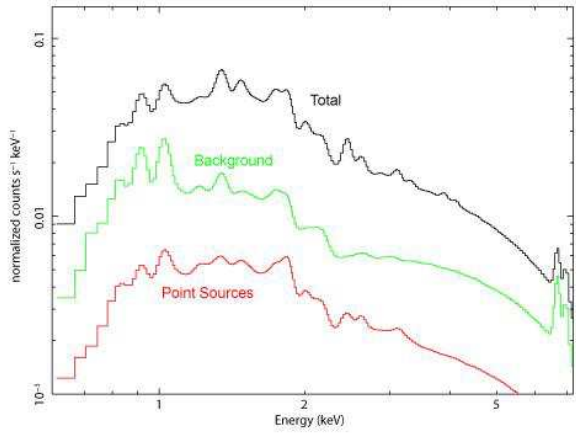}
    %%% \FigureFile(width,height){filename}
  \end{center}
  \caption{Contributions of the point sources (red) and the X-ray
 background (green) to the FI spectrum shown in Fig.~\ref{fig:xis}
 (black).}\label{fig:cont}
\end{figure}

The results of the fits are shown in Figure~\ref{fig:xis} and
Table~\ref{tab:fit}. The temperatures of the thermal components for 3T
model are $kT_1$, $kT_2$, and $kT_3$. Their intrinsic fluxes in the
0.7--10~keV band are $f_1$, $f_2$, and $f_3$, respectively. We assume
that the X-ray emission comes from the entire XIS field
($17\arcmin.8\times 17\arcmin.8$), and we make up the X-rays from the
masked regions with those from the surrounding regions. For 2TP model,
$kT_3$ and $f_3$ are replaced by the photon index ($\Gamma$) and the
non-thermal flux ($f_{\rm NT}$), respectively. The absorption $N_{H1}$
is the one for the lower temperature component ($kT_1$), and $N_{H2}$ is
the one for the higher. The errors in Table~\ref{tab:fit} include both
the statistical errors and the systematic errors from the uncertainties
about the GRXE and the CXB (section~\ref{sec:bg}). The low metal
abundances compared with the solar values are similar to those of the
Carina Nebula, which also contains massive stars \citep{ham07}. It is to
be noted that the contributing ratio of flux of the gas component, the
leaked stellar emission, and the X-ray background to the total spectrum
excluding the NXB in the 2--10~keV band is 43:14:43 (see
Figure~\ref{fig:cont}).

\citet{naz08} monitored bright stars in Westerlund~2 and found that
X-ray luminosities of several of the brightest stars vary about 30\%.
However, the sum of the luminosities of the 9 brightest stars in
Westerlund~2 is only $\sim 30$\% of the total stellar luminosity of
Westerlund~2. Moreover, it is unlikely that the flux variation of the
stars synchronizes. Considering the fact that only 14\% of the diffuse
X-ray emission around Westerlund~2 is attributed to the leaked stellar
emission, we can ignore the flux variation of the stars in the spectral
analysis.

\begin{table}
  \caption{Best-fit parameters for the gas component}\label{tab:fit}
  \begin{center}
    \begin{tabular}{lcc}
  \hline              
  Parameter & 3T & 2TP \\ 
      \hline
$N_{H1}$ ($10^{22}\rm\: cm^{-2}$)  
                & $0.96_{-0.37}^{+0.16}$ & $1.1_{-0.5}^{+0.0}$     \\
$k T_1$ (keV)   & $0.12_{-0.01}^{0.04}$  & $0.12_{-0.01}^{+0.04}$  \\
$N_{H2}$ ($10^{22}\rm\: cm^{-2}$)  
                & $1.1_{-0.3}^{+0.4}$    & $1.4_{-0.4}^{+0.2}$     \\
$k T_2$ (keV)   & $0.88_{-0.15}^{+0.21}$ & $1.0_{-0.1}^{+0.3}$     \\
$k T_3$ (keV)   & $4.2_{-1.2}^{+2.1}$    & .....                   \\
$\Gamma$        & .....                  & $2.2_{-0.4}^{+0.4}$     \\
C, N, O (solar) & $0.24_{-0.20}^{+0.71}$ & $0.57_{-0.43}^{+1.2}$   \\
Ne (solar)      & $0.04_{-0.04}^{+0.09}$ & $0.07_{-0.07}^{+0.17}$  \\
Mg (solar)      & $0.46_{-0.17}^{+0.32}$ & $0.77_{-0.42}^{+0.98}$  \\
Si (solar)      & $0.35_{-0.12}^{+0.20}$ & $0.38_{-0.19}^{+0.47}$  \\
S  (solar)      & $0.86_{-0.32}^{+0.43}$ & $0.78_{-0.31}^{+1.2}$   \\
Fe, Ni (solar)  & $0.0_{-0.0}^{+0.13}$   & $0.0_{-0.0}^{+0.33}$    \\
$f_1$ ($\rm ergs\: cm^{-2}\: s^{-1}$)   
                & $8.7_{-1.7}^{+0.0}\times 10^{-12}$   & $1.3_{-0.4}^{+0.3}\times 10^{-11}$    \\
$f_2$ ($\rm ergs\: cm^{-2}\: s^{-1}$)   
                & $2.2_{-0.0}^{+0.4}\times 10^{-12}$   & $3.0_{-0.2}^{+0.3}\times 10^{-12}$    \\
$f_3$ ($\rm ergs\: cm^{-2}\: s^{-1}$)   
                & $4.6_{-0.8}^{+0.5}\times 10^{-12}$   & .....  \\
$f_{\rm NT}$ ($\rm ergs\: cm^{-2}\: s^{-1}$)   
                & .....  & $5.4_{-1.4}^{+0.7}\times 10^{-12}$  \\
$\chi^2$/d.o.f. & 1719.56/1725           & 1729.06/1725            \\
\hline
    \end{tabular}
  \end{center}
\end{table}

\section{Discussion}
\label{sec:dicuss}

\subsection{Upper limit of Non-thermal Flux}

We found that the spectrum of the gas component of Westerlund~2 can be
represented by three thermal models (3T) or two thermal models and one
power-law model (2TP). Since the power-law component in 2TP model can be
replaced by a thermal component, the power-law component when the
contributions of the GRXE and the CXB to the total spectrum are minimum
gives the upper limit of non-thermal X-ray flux from the
cluster. Assuming that the non-thermal emission comes from the entire
XIS field, the upper limit of the flux is $f_{\rm NT}<2.6\times
10^{-12}\rm\: ergs\: cm^{-2}\: s^{-1}$ (0.7--2~keV) and $6.1\times
10^{-12}\rm\: ergs\: cm^{-2}\: s^{-1}$ (0.7--10~keV). In this
estimation, we compensate for the X-rays from the masked regions with
those from the surrounding regions.

Although the photon counts are small for detailed spectral analysis, we
found that X-ray surface brightness of a region that is $>8'$ away from
the cluster in the XIS field (the center of the circle is shifted from
the cluster center according to the surface brightness contours) is
comparable to the one estimated from the blank region. Thus, the
emission from the outside of the XIS field does not much contribute to
the total diffuse emission from Westerlund~2.

We also analyzed the HXD data of Suzaku. Adopting a systematic
uncertainty of 5\% for the NXB, we found that the upper limit of the
X-ray flux from Westerlund~2 in the 15--40~keV band is $5\times
10^{-12}\rm\: ergs\: cm^{-2}\: s^{-1}$, excluding the CXB and the
GRXE. We assumed that the spectrum of the CXB is represented by a
power-law model multiplied by a HIGHECUT model in XSPEC. We adopted the
power-law index of 1.29, the power-law normalization of $9.412\times
10^{-3}\rm\: photons\; cm^{-2}\: s^{-1}\: FOV^{-1} keV^{-1}$, and the
e-folding energy of 40~keV, following a standard
model\footnote{http://heasarc.gsfc.nasa.gov/docs/suzaku/analysis/pin\_cxb.html}. The
spectrum of the GRXE is extrapolated from the one in the soft X-ray
band. The upper limit of the flux from Westerlund~2 can be compared
with the extrapolated values of 3T and 2TP models ($\lesssim 4\times
10^{-12}\rm\: ergs\: cm^{-2}\: s^{-1}$). Considering that the HXD has a
wider field ($34'\times 34'$) than the XIS, this result indicates that
it is unlikely that strong hard X-ray emission extends beyond the XIS
field.

\subsection{Comparison with Previous Observations}

\begin{figure}
  \begin{center}
    \FigureFile(80mm,80mm){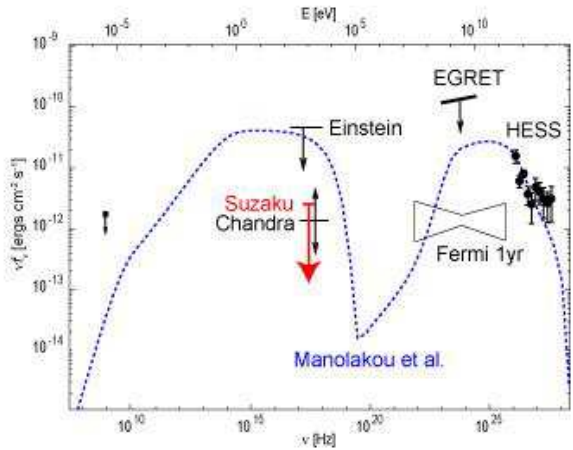}
    %%% \FigureFile(width,height){filename}
  \end{center}
  \caption{Multi-wavelength measurements of Westerlund~2 (radio:
 \cite{whi97}; Einstein: \cite{gol87}; Chandra: \cite{tow05}; Suzaku:
 this study [0.7--2~keV]; EGRET; \cite{har99}; H.E.S.S.:
 \cite{aha07}). Fermi one year flux sensitivity is shown by the thin
 solid line\footnotemark[3]. A theoretical prediction of \citet{man07}
 is shown by the dotted line (their model of
 $t=10^5$~yr).}\label{fig:flux}
\end{figure}

In Figure~\ref{fig:flux}, we present multi-wavelength measurements of
Westerlund~2. The radio flux is considered as an upper limit, because it
could only partially include the non-thermal radio flux produced by the
energetic electrons \citep{whi97}. The Einstein observations probably
overestimate the diffuse flux, because the limited spatial resolution
may not allow the subtraction of the point sources well enough
\citep{gol87}. The flux obtained with EGRET may also need to be regarded
as an upper limit, because of its low spatial resolution ($\sim
1\degree$).

\footnotetext[3]{http://www-glast.slac.stanford.edu/software/IS/glast\_lat\_performance.htm}

With Chandra, \citet{tow05} found that the diffuse X-ray emission from
Westerlund~2 can be represented by three thermal components ($kT=0.1$,
0.8, and 3.1~keV), which is consistent with our 3T model
(Table~\ref{tab:fit}). \citet{tow05} indicated that the hottest
component can be replaced with a power-law component
($\Gamma=2.3$). This photon index is also consistent with that in
Table~\ref{tab:fit}.  \citet{tow05} estimated that the flux of the
power-law component is $1.4\times 10^{-12}\rm\: ergs\: cm^{-2}\: s^{-1}$
(0.7--10~keV). The difference from our best-fit value ($f_{\rm
NT}=5.4\times 10^{-12}\rm\: ergs\: cm^{-2}\: s^{-1}$; see
Table~\ref{tab:fit}) could be due to the difference of the regions
investigated, and the backgrounds adopted. \citet{tow05} focused on a
narrower region ($\lesssim 5'$ from the center of Westerlund~2), and
they took a background in the same Chandra field. Thus, the flux should
be regarded as an lower limit. On the other hand, since the power-law
component can be replaced by a thermal component, the flux could also be
regarded as an upper limit (Figure~\ref{fig:flux}).

With the high-quality of the Suzaku spectrum and the wide region we
covered, we think that we have obtained the strictest upper limit of the
diffuse non-thermal emission from Westerlund~2 so far.

\subsection{Comparison with Theoretical Models of Particle Acceleration}

Since Westerlund~2 has more than ten O~stars and two Wolf-Rayet (WR)
stars \citep{rau07}, the strong winds from them may currently
accelerate particles, which may be responsible for the TeV gamma-ray
emission from the cluster. Considering the source extention, it is
unlikely that a single star produces the gamma-ray emission
\citep{aha07}. \citet{bed07} and \citet{man07} calculated broad band
spectra of Westerlund~2; they chose parameters so that the spectra are
consistent with previous gamma-ray and X-ray observations. They
predicted that synchrotron emission from accelerated electrons could be
observed in the X-ray band. Assuming that the distance to Westerlund~2
is 8~kpc and particles are injected uniformly during the lifetime of a
WR star, \citet{bed07} predicted that the X-ray flux would be $1.3\times
10^{-11}\rm\: ergs\: cm^{-2}\: s^{-1}$ at $\sim 1$~keV. Assuming that
the distance to the cluster is 2.8~kpc, \citet{man07} predicted that the
X-ray flux would be $\sim 2$--$3\times 10^{-11}\rm\: ergs\: cm^{-2}\:
s^{-1}$ at $\sim 1$~keV, depending on the injection time-scale. On the
other hand, our new Suzaku results show that the upper limit of the
non-thermal emission is $f_{\rm NT}<2.6\times 10^{-12}\rm\: ergs\:
cm^{-2}\: s^{-1}$ (0.7--2~keV), which is much smaller than the above
predictions (Figure~\ref{fig:flux}). It would be difficult to adjust the
parameters of the theoretical models to match both the new X-ray
constraint and the previous gamma-ray observations, because the models
must satisfy the relatively large ratio of gamma-ray to X-ray energy
flux.

Recently, \citet{fuk09} discovered jet and arc-like molecular feature
around Westerlund~2. In particular, the latter suggests a past stellar
explosion (or past stellar explosions) in the cluster. Based on this
observation, we consider particle acceleration associated with the
explosion. 

The total mass of the molecular gas around Westerlund~2 is $\sim 2\times
10^5\: M_\odot$ and the size is $\sim 30$~pc \citep{fur09}. If the
molecular cloud before the explosion of the star had almost the same
mass and size, the average proton number density is $\sim 100\rm\:
cm^{-3}$. If the star explodes $\sim 5\times 10^5$~yrs ago in the
molecular cloud with the kinetic energy of $10^{51}$~erg, the current
velocity and radius of the shock wave is $\sim 16\rm\: km\: s^{-1}$ and
$\sim 40$~kpc, respectively (see equation~2 in \cite{yam06}. The radius
can be obtained by the integration of the velocity). The velocity is
comparable to the internal velocity of the molecular gas, and the size
is comparable to that of the arc ($\sim 30$~pc) observed around
Westerlund~2 \citep{fuk09}.  Although the shock may still be expanding,
it is unlikely that particles are accelerated at the shock at present
because of the small velocity. However, particles might be accelerated
in the past when the velocity of the shock was large.  If adiabatic
energy loss did not much affect the particle energy, the accelerated
protons should not have lost their energy because of their long cooling
time. Thus, the protons accelerated in the past may be producing the TeV
gamma-rays through $pp$-interactions in the surrounding molecular
gas. On the other hand, accelerated electrons should have lost their
energy through synchrotron emission. Therefore, the current ratio of
gamma-ray to X-ray energy flux should be large, because the electrons
are the source of the X-ray flux through synchrotron emission
\citep{yam06}. This is consistent with the observations of Westerlund~2;
the observed ratio of the 1--10~TeV energy flux to the 2--10~keV energy
flux is $R_{\rm TeV/X}>2.7$, which actually indicates that the stellar
explosion did not happen recently \citep{yam06}. Such stellar explosions
in dense clouds could be the sources of unidentified TeV sources and the
so-called PAMELA anomaly \citep{bam09,iok09,fuj09}.

The mass of the star responsible for the explosion could be extremely
large, because stars in an open cluster form almost simultaneously and
massive stars with $\sim 80\: M_\odot$ still survive in Westerlund~2
\citep{rau05}. In general, such massive stars produce ejecta with a
large $\alpha$-element to iron abundance ratio \citep{kob06}. This trend
is consistent with the metal abundances shown in Table~\ref{tab:fit},
although the contribution from stellar winds from surviving stars must
be considered for detailed analysis. If the thermal gas in 3T model is
uniformly distributed within $8'$ from the cluster center, the total gas
mass is $890_{-290}^{+160}\: M_\odot$. From the metal abundances in
Table~\ref{tab:fit}, the masses of Si and S contained in the gas are
estimated to be $\sim 0.1$--$0.6\: M_\odot$ and $\sim 0.1$--$0.7\:
M_\odot$, respectively. Since a star with a mass of $\gtrsim 40\:
M_\odot$ produces $\sim 0.3\: M_\odot$ of Si and $\sim 0.1\: M_\odot$ of
S \citep{kob06}, explosions of a few massive stars are enough to produce
observed Si and S, even if metals that originate from sources other than
the stellar explosions are not considered. Since some of the stars with
masses of $\gtrsim 30\: M_\odot$ explode as hypernovae
\citep{gal98,iwa98}, the one exploded in Westerlund~2 might be such a
hypernova and might trigger a gamma-ray burst. However, because of the
large errors for the metal abundances in Table~\ref{tab:fit} and the
lack of the knowledge about the mass of the progenitor star, we cannot
firmly determine whether the progenitor star is a supernova or a
hypernova from the metal abundances \citep{kob06}.

It is to be noted that pure hadronic models could not account for the
GeV gamma-ray radiation detected by EGRET at the position of
Westerlund~2 \citep{har99,bed07}. In the near future, Fermi will reveal
whether the GeV gamma-ray source is related to the TeV gamma-ray source
in Westerlund~2 and by what mechanism the gamma-ray radiation is
produced (Figure~\ref{fig:flux}).

\section{Conclusion}
\label{sec:conc}

We observed the young open cluster Westerlund~2 with Suzaku X-ray
satellite.  We found that diffuse X-ray emission extends to $\sim 8'$
from the cluster center. We analyze the spectrum considering the
contamination from point sources using Chandra data. We found that the
diffuse emission consists of three thermal components ($kT\sim 0.1$,
0.9, and 4~keV) or two thermal components ($kT\sim 0.1$ and 1.0~keV) and
one non-thermal component ($\Gamma\sim 2.2$). The upper limit of the 
non-thermal energy flux is smaller than the TeV gamma-ray energy flux
observed with H.E.S.S. The abundances of $\alpha$-elements are
relatively high in comparison with that of iron.

The relatively high X-ray to gamma-ray energy flux ratio suggests that
the gamma-ray emission is attributed to protons that were accelerated in
the past, because the cooling time of the high-energy protons is much
longer than that of high-energy electrons, and the electrons that are
responsible for synchrotron X-ray emission may have cooled. Considering
the structures in molecular gas surrounding the cluster, protons would
have been accelerated at a shock created through a stellar explosion
that happened $\sim 10^5$--$10^6$ yrs ago. Since extremely massive stars
($\sim 80\: M_\odot$) still survive in this cluster, the star
responsible for the explosion would also have been very massive. The
high abundances of $\alpha$-elements compared with iron may support this
idea. Thus, the progenitor star could have exploded as a hypernova
rather than as a normal supernova.

\vspace{3mm}

We wish to thank M.~Tsujimoto for providing us the Chandra data. We also
thank K.~Hamaguchi, and Y.~Hyodo for useful discussion. This work was
supported in part by a Grant-in-aid from the Ministry of Education,
Culture, Sports, Science, and Technology (MEXT) of Japan, No. 20540269
(Y.~F.).

%%%


\begin{thebibliography}{}

\bibitem[Abdo et al.(2007)]{abd07} Abdo, A.~A., et al.\
2007, \apjl, 658, L33

\bibitem[Aharonian et 
al.(2002)]{aha02} Aharonian, F., et al.\ 2002, \aap, 393,
	  L37

\bibitem[Aharonian et al.(2004)]{aha04} Aharonian, F.~A., et 
al.\ 2004, \nat, 432, 75 

\bibitem[Aharonian et 
al.(2007)]{aha07} Aharonian, F., et al.\ 2007, \aap, 467, 1075 

\bibitem[Anchordoqui et al.(2007)]{anc07} Anchordoqui, L.~A., 
Beacom, J.~F., Goldberg, H., Palomares-Ruiz, S.,
\& Weiler, T.~J.\ 2007, Physical Review Letters, 98, 121101 

\bibitem[Ascenso et 
al.(2007)]{asc07} Ascenso, J., Alves, J., Beletsky, Y., \& Lago,
		M.~T.~V.~T.\ 2007, \aap, 466, 137

\bibitem[Bamba et al.(2009)]{bam09} Bamba, A., Yamazaki, R., 
Kohri, K., Matsumoto, H., Wagner, S., P{\"u}hlhofer, G., 
\& Kosack, K.\ 2009, \apj, 691, 1854

\bibitem[Bednarek(2003)]{bed03} Bednarek, W.\ 2003, \mnras, 
345, 847

\bibitem[Bednarek(2007)]{bed07} Bednarek, W.\ 2007, \mnras, 
382, 367 

\bibitem[Belloni 
\& Mereghetti(1994)]{bel94} Belloni, T., \& Mereghetti, S.\ 1994, \aap,
		286, 935 

\bibitem[Broos et al.(2002)]{Broos02} Broos, P. S., Townsley, L. K.,
Getman, K., \& Bauer, F. E.\ 2002, ACIS Extract, An ACIS Point Source
Extraction Package (University Park: The Pennsylvania State Univ.)
http://www.astro.psu.edu/xray/docs/TARA/ae\_users\_guide.html

\bibitem[Dame(2007)]{dam07} Dame, T.~M.\ 2007, \apjl, 665, 
L163 

\bibitem[Dickey 
\& Lockman(1990)]{dic90} Dickey, J.~M., \& Lockman, F.~J.\ 1990, \araa,
		28, 215

\bibitem[Ebisawa et al.(2005)]{ebi05} Ebisawa, K., et al.\ 
2005, \apj, 635, 214 

\bibitem[Ezoe et al.(2009)]{ezo09} Ezoe, Y., Hamaguchi, K., 
Gruendl, R.~A., Chu, Y.-H., Petre, R., 
\& Corcoran, M.~F.\ 2009, \pasj, 61, 123

\bibitem[Fujita et al.(2009)]{fuj09} Fujita, Y., Kohri, K., 
Yamazaki, R., \& Ioka, K.\ 2009, arXiv:0903.5298 
%%CITATION = ARXIV:0903.5298;%%

\bibitem[Fukui et al.(2009)]{fuk09} Fukui, Y.,
Furukawa, N., Dame, T.M., Dawson, J.R., Yamamoto, H., Rowell, G.P., 
Aharonian, F., Hofmann, W., de O{\~n}a Wilhelmi, E., Minamidani, T., 
Kawamura, A., Mizuno, N., Onishi, T., Mizuno, A., and Nagataki, S.: 2009, 
arXiv:0903.5340. 

\bibitem[Furukawa et al.(2009)]{fur09} Furukawa, N., Dawson, 
J.~R., Ohama, A., Kawamura, A., Mizuno, N., Onishi, T., 
\& Fukui, Y.\ 2009, \apjl, 696, L115 

\bibitem[Galama et al.(1998)]{gal98} Galama, T.~J., et al.\ 
1998, \nat, 395, 670 

\bibitem[Getman et al.(2005)]{Getman05b} Getman, K.~V., et al.\
2005, \apjs, 160, 319

\bibitem[Giovannelli et al.(1996)]{gio96} Giovannelli, F., 
Bednarek, W., 
\& Karakula, S.\ 1996, Journal of Physics G Nuclear Physics, 22, 1223 

\bibitem[Goldwurm et al.(1987)]{gol87} Goldwurm, A., Caraveo, 
P.~A., \& Bignami, G.~F.\ 1987, \apj, 322, 349 

\bibitem[Hamaguchi et al.(2007)]{ham07} Hamaguchi, K., et 
al.\ 2007, \pasj, 59, 151 

\bibitem[Hartman et al.(1999)]{har99} Hartman, R.~C., et al.\ 
1999, \apjs, 123, 79 

\bibitem[Hertz 
\& Grindlay(1984)]{her84} Hertz, P., \& Grindlay, J.~E.\ 1984, \apj,
		278, 137

\bibitem[Hyodo et al.(2008)]{hyo08} Hyodo, Y., Tsujimoto, M., 
Hamaguchi, K., Koyama, K., Kitamoto, S., Maeda, Y., Tsuboi, Y., 
\& Ezoe, Y.\ 2008, \pasj, 60, 85 

\bibitem[Ioka 
\& Meszaros(2009)]{iok09} Ioka, K., \& Meszaros, P.\ 2009,
		arXiv:0901.0744
%%CITATION = ARXIV:0901.0744;%%

\bibitem[Ishisaki et al.(2007)]{ish07} Ishisaki, Y., et al.\ 
2007, \pasj, 59, 113 

\bibitem[Iwamoto et al.(1998)]{iwa98} Iwamoto, K., et al.\ 
1998, \nat, 395, 672 

\bibitem[Kalberla et 
al.(2005)]{kal05} Kalberla, P.~M.~W., Burton, W.~B., Hartmann, D.,
		Arnal, E.~M., Bajaja, E., Morras, R., P{\"o}ppel,
		W.~G.~L.\ 2005, \aap, 440, 775

\bibitem[Koyama et al.(2007)]{koy07} Koyama, K., et al.\ 
2007, \pasj, 59, 23 

\bibitem[Koyama et al.(1995)]{koy95} Koyama, K., Petre, R., 
Gotthelf, E.~V., Hwang, U., Matsuura, M., Ozaki, M., 
\& Holt, S.~S.\ 1995, \nat, 378, 255 

\bibitem[Kobayashi et al.(2006)]{kob06} Kobayashi, C., Umeda, 
H., Nomoto, K., Tominaga, N., \& Ohkubo, T.\ 2006, \apj, 653, 1145 

\bibitem[Kushino et al.(2002)]{kus02} Kushino, A., Ishisaki, 
Y., Morita, U., Yamasaki, N.~Y., Ishida, M., Ohashi, T., 
\& Ueda, Y.\ 2002, \pasj, 54, 327 

\bibitem[Manolakou et 
al.(2007)]{man07} Manolakou, K., Horns, D., \& Kirk, J.~G.\ 2007, \aap,
		474, 689

\bibitem[Mitsuda et al.(2007)]{mit07} Mitsuda, K., et al.\ 
2007, \pasj, 59, 1 

\bibitem[Naz{\'e} et 
al.(2008)]{naz08} Naz{\'e}, Y., Rauw, G., \& Manfroid, J.\ 2008, \aap,
		483, 171

\bibitem[Piatti et 
al.(1998)]{pia98} Piatti, A.~E., Bica, E., \& Claria, J.~J.\ 1998,
		\aaps, 127, 423

\bibitem[Rauw et al.(2005)]{rau05} Rauw, G., et al.\ 2005, \aap, 432,
		985

\bibitem[Rauw et 
al.(2007)]{rau07} Rauw, G., Manfroid, J., Gosset, E., Naz{\'e}, Y.,
		Sana, H., De Becker, M., Foellmi, C., \& Moffat,
		A.~F.~J.\ 2007, \aap, 463, 981

\bibitem[Revnivtsev et al.(2009)]{rev09} Revnivtsev, M., 
Sazonov, S., Churazov, E., Forman, W., Vikhlinin, A., 
\& Sunyaev, R.\ 2009, \nat, 458, 1142 

\bibitem[Revnivtsev et 
al.(2006)]{rev06} Revnivtsev, M., Sazonov, S., Gilfanov, M., Churazov,
		E., \& Sunyaev, R.\ 2006, \aap, 452, 169

\bibitem[Snowden et al.(1998)]{sno98} Snowden, S.~L., Egger, 
R., Finkbeiner, D.~P., Freyberg, M.~J., 
\& Plucinsky, P.~P.\ 1998, \apj, 493, 715 

\bibitem[Tawa et al.(2008)]{taw08} Tawa, N., et al.\ 2008, 
\pasj, 60, 11 

\bibitem[Torres et al.(2004)]{tor04} Torres, D.~F., 
Domingo-Santamar{\'{\i}}a, E., \& Romero, G.~E.\ 2004, \apjl, 601, L75 

\bibitem[Townsley et al.(2003)]{TFM03} Townsley, L.~K.,
Feigelson, E.~D., Montmerle, T., Broos, P.~S., Chu, Y.-H., \& Garmire,
G.~P.\ 2003, \apj, 593, 874

\bibitem[Townsley et al.(2005)]{tow05} Townsley, L.~K.,
Feigelson, E.~D., Montmerle, T., Broos, P.~S., Chu, Y.-H., Garmire,
G.~P.\& Getman K.\ 2005, in Sjouwerman L. O., Dyer K. K., eds, Proc.
X-ray and Radio Connections, Santa Fe, http://www.aoc.nrao.edu/
events/xraydio/meetingcont-/3.4\_townsley.pdf

\bibitem[Tsujimoto et al.(2007)]{tsu07} Tsujimoto, M., et 
al.\ 2007, \apj, 665, 719 

\bibitem[Whiteoak 
\& Uchida(1997)]{whi97} Whiteoak, J.~B.~Z., \& Uchida, K.~I.\ 1997,
		\aap, 317, 563

\bibitem[Yamazaki et al.(2006)]{yam06} Yamazaki, R., Kohri, 
K., Bamba, A., Yoshida, T., Tsuribe, T., 
\& Takahara, F.\ 2006, \mnras, 371, 1975 
%%CITATION = MNRAA,371,1975;%%

\end{thebibliography}
\end{document}